\documentclass[a4paper,prb,aps,epsfig,twocolumn,showpacs,preprintnumbers,superscriptaddress,float,amsmath,amssymb]{revtex4}

\usepackage{graphicx}
\usepackage{dcolumn}
\usepackage{bm}

\usepackage{color}
\usepackage[normalem]{ulem}
\usepackage{graphicx}
\usepackage{pgfplots}
\usetikzlibrary{arrows.meta}
\pgfplotsset{compat=1.16}

\begin{document}

\title{The fate of topological frustration in quantum spin ladders and generalizations}
\date{\today}

\author{Po-Wei Lo}
\affiliation{Laboratory of Atomic and Solid State Physics,
Cornell University, Ithaca, NY, 14853}
\author{Michael J Lawler}
\affiliation{Laboratory of Atomic and Solid State Physics,
Cornell University, Ithaca, NY, 14853}
\affiliation{Department of Physics, Applied Physics and Astronomy, Binghamton University, Binghamton, New York 13902}

\begin{abstract}
Topological frustration (or topological mechanics) is the existence of classical zero modes that are robust to many but not all distortions of the Hamiltonian. It arises naturally from locality in systems whose interactions form a set of constraints such as in geometrically frustrated magnets and balls and springs metamaterials. For a magnet whose classical limit exhibits topological frustration, an important question is what happens to this topology when the degrees of freedom are quantized and whether such frustration could lead to exotic quantum phases of matter like a spin liquid. We answer these questions for a geometrically frustrated spin ladder model. It has the feature of having infinitely many conserved quantities that aid the solution. We find classical zero modes all get lifted by quantum fluctuations and the system is left with a unique rung singlet ground state---a trivial quantum spin liquid. Moreover, we find low-energy eigenstates corresponding to known symmetry protected topological (SPT) ground states, and a special role of $SU(2)$ symmetry, that it demands the existence of extra dimensions of classical zero modes---the phenomena we call symmetry-enriched topological frustration (SETF).
These results suggest small violations of the conservation laws in the nearly SETF regime could lead to quantum scars. 
We further study a two-dimensional bilayer triangular lattice model and find a similar SETF phenomena which also leads to suppressed low-energy topological eigenstates in the quantum regime.
These results suggest that in the absence of magnetic order, classical topological frustration manifests at finite spin as asymptotically low energy modes with support for exotic quantum phenomena.



\end{abstract}

\maketitle

\section{introduction}

Geometric frustration in physics brings many exotic phenomena. 
Spin lattices such as triangular, kagome, and pyrochlore lattices 
may possess novel quantum phenomena\cite{PhysRevB.50.10048,PhysRevLett.82.3899,PhysRevB.46.14201,PhysRevLett.80.2933,PhysRevB.84.064505} and novel classical phenomena arising from a large degeneracy of classical ground states such as 
spin origami\cite{PhysRevLett.70.3812,PhysRevB.47.15342,PhysRevLett.121.177201}. These classical phenomena seem at first sight disconnected from the quantum phenomena but perhaps not.
If we could solve some of these frustrated spin systems in the large-but-finite spin regime, what kinds of quantum phenomena would be revealed?

In many cases, it is known that the degeneracy of classical ground states is lifted by quantum fluctuations and the ground state becomes magnetically ordered in two or three-dimensional space. This order-by-disorder phenomenon has been established by performing perturbative expansions or spin-wave approximation approaches in the large spin $S$ limit. We know for example that in Heisenberg antiferromagnets on a triangular, square, or kagome lattice, order-by-disorder occurs\cite{PhysRevB.46.11137,PhysRevB.95.014425,PhysRevB.56.2521}. However, a magnetically ordered phase may not always be the fate of a frustrated spin system. Taking into account higher-order corrections in $1/S$, it has been shown that a magnetic order ground state is not easily established and may even be absent in some frustrated spin systems, including some square lattice models\cite{PhysRevB.38.9335,PhysRevB.101.214404} and the pyrochlore Heisenberg antiferromagnet\cite{PhysRevLett.79.2554,Hizi_2007}. 
In the cases where an ordered phase is not the fate of frustration, can frustration support exotic quantum phenomena? If so, what kinds of frustration can support these phenomena? 


We consider the case of geometric frustration and whether it can support exotic quantum phenomena at large-but-finite $S$. In particular, we are focused on the case where the Hamiltonian of a geometrically frustrated spin system can be written as a classical frustration-free form. In this case, the classical ground state can be understood as zero modes of a constrained problem. The zero modes of classical spins then obey Moessner-Chalker-Maxwell counting\cite{Maxwell1864} and form mechanical analogs of topological mechanics\cite{Kane2013} like spin origami\cite{PhysRevLett.121.177201}. We call this phenomenon ``topological frustration'' and wonder whether it supports topological quantum states, quantum spin liquids, or other exotic quantum phenomena. 

Our approach is to study the connections between topological frustration and quantum magnets in one-dimensional space where an ordered phase is naturally prevented. We do so by studying quantum spin ladders\cite{PhysRevB.57.11439} in a special regime where the classical spins exhibit geometric frustration on each plaquette similar to classical spins on each tetrahedron of pyrochlore Heisenberg antiferromagnets. A key distinction from the pyrochlore case beyond dimension is the additional existence of infinitely many conservation laws owing to the existence of symmetry we call "staggered swap" symmetry. 
Using these conservation laws, we show that the fate of classical topological frustration in the quantum regime is to emerge as asymptotically-in-$S$ low energy low-entanglement eigenstates. These eigenstates violate the eigenstate thermalization hypothesis, have area law entanglement, and correspond to known symmetry-protected topological (SPT) ground states enabled by the staggered swap symmetry. We further identify a special role of $SU(2)$ symmetry, that it demands the existence of extra dimensions of classical zero modes the phenomena we call symmetry-enriched topological frustration (SETF). We conclude with a discussion of a) how small violations of the special symmetries used to obtain results in this paper would likely lead to quantum scars, b) the model generalizes to higher dimensions proposing a bilayer triangular lattice model which shares many similar properties with our quantum spin ladders, and c) a discussion of why we think these results suggest tensor network methods are a powerful approach to the study of large-$S$ antiferromagnets. 

\section{Geometrically frustrated spin ladders}

\begin{figure}
  \includegraphics[width=8cm]{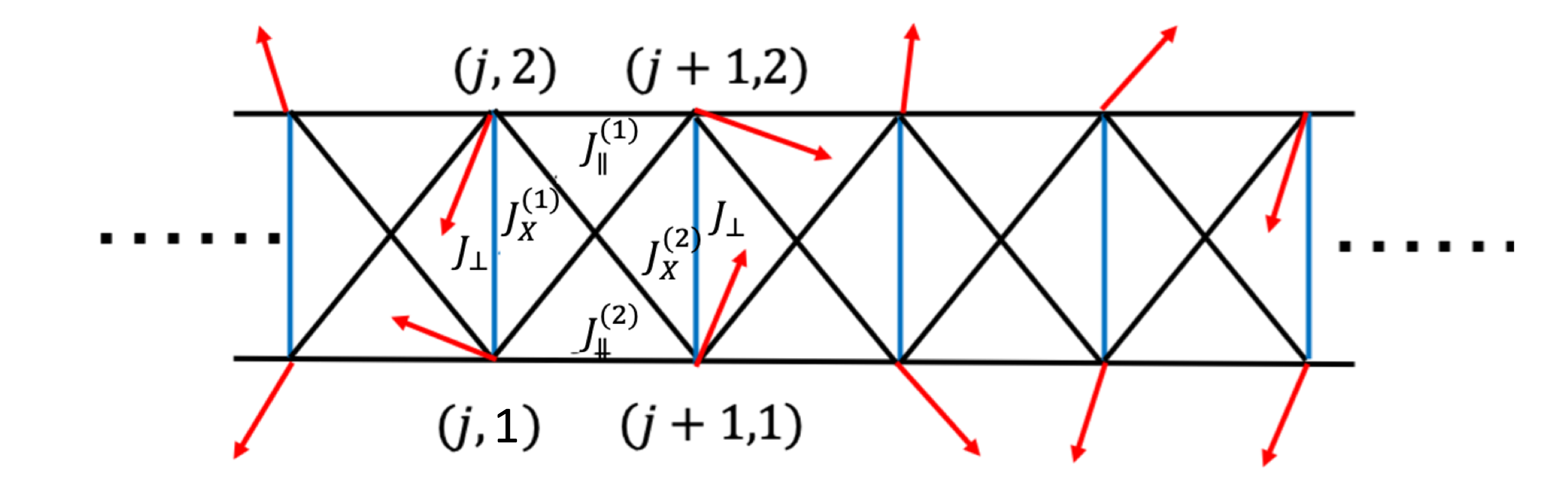}
  \caption{The frustrated spin ladder model}
  \label{fig:ladder}
\end{figure}

We start with a frustrated spin ladder model (Fig.\ref{fig:ladder}) which possesses classical frustration, local conserved quantities, or both in some regimes. The Hamiltonian is written as
\begin{equation}\label{eq::ladder}
\begin{split}
    H_{ladder}=\frac{J_{\perp}}{S(S+1)}\sum_{i}{\overrightarrow S_{i,1} \cdot \overrightarrow S_{i,2}}\\
    +\frac{J_{\parallel,1}}{S(S+1)}\sum_{i}{\overrightarrow S_{i,1} \cdot \overrightarrow S_{i+1,1}}\\
    +\frac{J_{\parallel,2}}{S(S+1)}\sum_{i}{\overrightarrow S_{i,2} \cdot \overrightarrow S_{i+1,2}}\\
    +\frac{J_{X,1}}{S(S+1)}\sum_{i}{\overrightarrow S_{i,1} \cdot \overrightarrow S_{i+1,2}}\\
    +\frac{J_{X,2}}{S(S+1)}\sum_{i}{\overrightarrow S_{i,2} \cdot \overrightarrow S_{i+1,1}}
\end{split}
\end{equation}
where $\overrightarrow S_{i,m}$ is the spin operator at the site $(i,m)$, and $J_{\perp}$, $J_{\parallel,1}$, $J_{\parallel,2}$, $J_{X,1}$, and $J_{X,2}$ are the antiferromagnetic coupling strengths depicted in Fig.\ref{fig:ladder}. We divide the antiferromagnetic coupling strength by $(S+1)S$ so that as spin increasing we only increase the number of degrees of freedom but keep unit length for the spin.


This spin ladder model looks complex but has been studied for the spin-half case in many different regimes\cite{PhysRevB.48.10653,PhysRevB.52.12485,PhysRevB.57.11439}. In general, it relies on numerical simulations to find the ground state\cite{PhysRevB.57.11439} except for some regimes which can be studied analytically\cite{PhysRevB.48.10653,PhysRevB.52.12485}. For example, in the limit $J_{\perp}\gg J_{\parallel,1},J_{\parallel,2},J_{X,1},J_{X,2}$, the ground state is a gapped rung singlet\cite{PhysRevB.48.10653}. Another well-controlled regime is the case where $J_1=J_{\parallel,1}=J_{\parallel,2}=J_{X,1}=J_{X,2}$\cite{PhysRevB.52.12485}. In this regime, the Hamiltonian has two competing phases. One of the phases is the rung singlet in which the coupling $J_{\perp}$ between two spins at the same rung dominates, and thus the system forms a singlet spin state at each rung. The other phase is the spin-one Haldane phase in which two spins at the same rung are aligned to the same direction forming an equivalently spin-one quasiparticle that couples to its two neighbors and behaves just as the Heisenberg spin-one chain. Despite the simple structure of the phase diagram, this well-controlled regime can be highly frustrated in the classical limit and thus a good candidate to study any connections between classical geometric frustration and quantum magnetism.

\section{Phase diagram}

To establish the connections, we generalize the spin ladder model to arbitrary spin $S$. First, in the classical $S\to\infty$ limit, the spin operator $\overrightarrow S_{i,m}$ is reduced to a three-dimensional vector. One can use Lagrange multipliers to fix the length of each spin, and then the ground states are obtained by minimizing the energy with respect to each spin component. Interestingly, there exists a special regime where the Hamiltonian can be written in a frustration-free form
\begin{equation}
\begin{split}
    H_{fru}=\frac{J}{S(S+1)}\sum_{i}(a_1\overrightarrow S_{i,1} + a_2 \overrightarrow S_{i,2}\\ 
    +a_3\overrightarrow S_{i+1,1}
    + a_4 \overrightarrow S_{i+1,2})^2
\end{split}
\label{eq:frustrated}
\end{equation}
which requires two conditions $J_{\parallel,1}J_{\parallel,2}=J_{X,1}J_{X,2}$ and $J_{\perp} \geq 2\sqrt{J_{\parallel,1}J_{\parallel,2}}$. The Hamiltonian $H_{fru}$ has large
ground state degeneracy, and thus the system is highly frustrated. In this type of highly frustrated regime, the classical ground state can be understood as zero modes of a constraint problem. In the spin ladder model, the configurations of zero modes can be obtained by sequentially add two spins on the $i$th rung that satisfy the constraint
\begin{equation}\label{eq:constraint}
    a_1\overrightarrow S_{i,1} + a_2\overrightarrow S_{i,2}+a_3\overrightarrow S_{i+1,1}+ a_4 \overrightarrow S_{i+1,2}=0.
\end{equation}

In the quantum finite $S$ regime, similarly to the spin-half case, only some regimes can be studied analytically. Especially, we are interested in the frustrated but well-controlled regime where $J_1=J_{\parallel,1}=J_{\parallel,2}=J_{X,1}=J_{X,2}$. In this regime, the ladder has global staggered swap symmetry: the Hamiltonian is invariant by swapping the two spins on all the even rungs or all the odd rungs. In this regime the Hamiltonian also has the local conservation law with the conserved quantum number $T_i(T_i+1)$ where $T_i$ is the total spin quantum number on a rung with spin operator defined as $\overrightarrow T_i=\overrightarrow S_{i,1}+ \overrightarrow S_{i,2}$. Thus, we can rewrite the Hamiltonian as
\begin{equation}
    H_{con}=\frac{J_{\perp}/2}{S(S+1)}\sum_{i}{\overrightarrow T_{i}^2}
    +\frac{J_{1}}{S(S+1)}\sum_{i}{\overrightarrow T_{i} \cdot \overrightarrow T_{i+1}} + const.
\end{equation}

For a given spin $S$, there are $2S+1$ competing phases, one for each value of spin representation $T_i$ including the rung singlet ($T_i=0$) and the well-known SPT Haldane phases from $T_i=1$ to $T_i= 2S$. However, only the rung singlet and the spin-$2S$ Haldane state can be the ground state depending on the ratio of the antiferromagnetic coupling strengths $J_{\perp}/J_1$. When $J_{\perp}$ dominates, two spins on the same rung form a spin-singlet. When the coupling $J_1$ between two neighbor rungs dominates, each rung forms a maximum spin-$2S$ quasiparticle which antiferromagnetically couples to its two neighbors, and the spin ladder model is equivalent to the spin-$2S$ Heisenberg chain.

\begin{figure}
  \includegraphics[width=8cm]{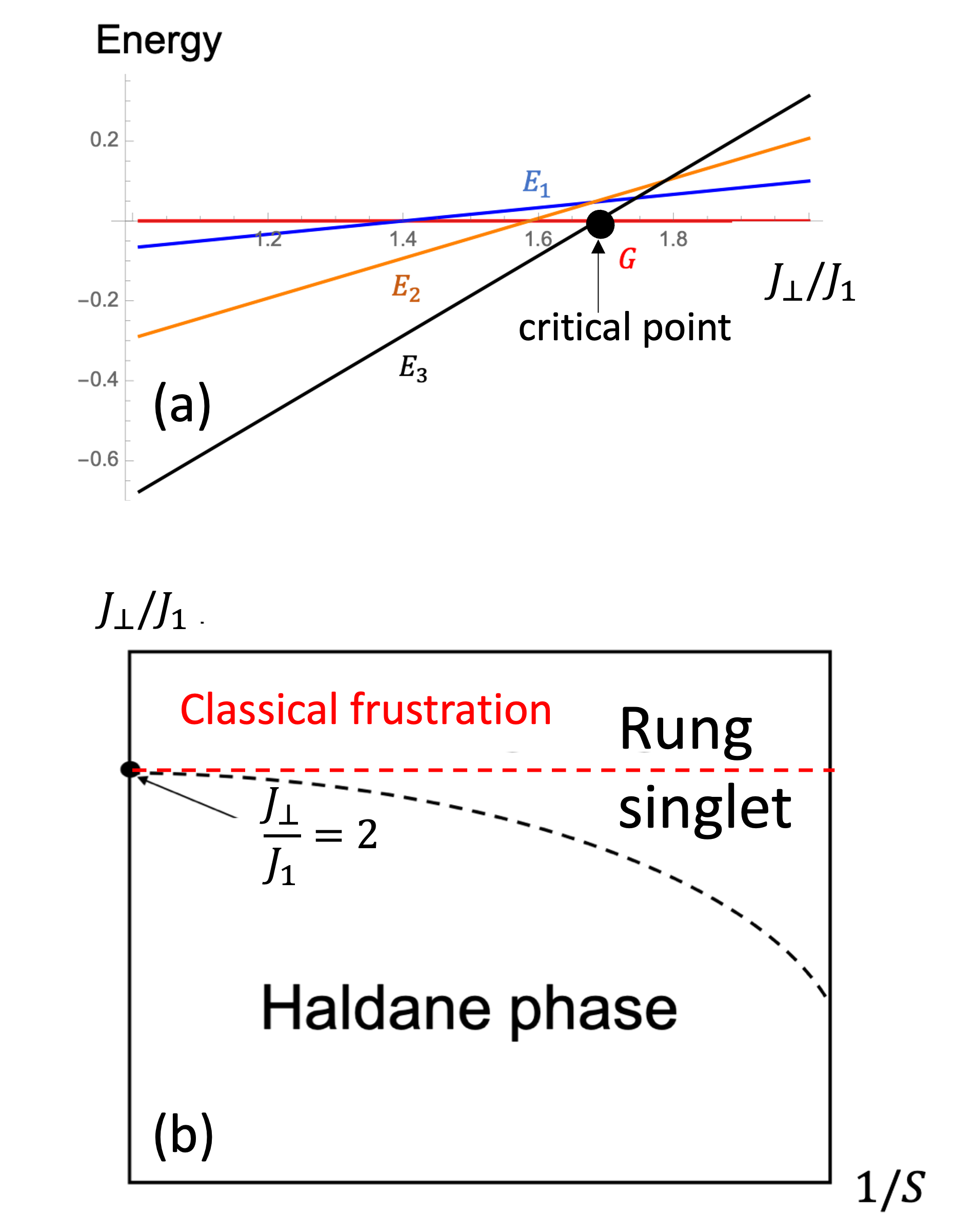}
  \caption{(a)The density-matrix renormalization group calculation for the eigenenergy of the spin-$3/2$ frustated spin ladder model ($50$ rungs) for different values of $J_{\perp}/J_1$. Here we fix $J_1=1$. $G$: the rung singlet; $E_i$: the spin-$i$ Haldane state.(b)A schematic phase diagram of the frustrated spin ladder model}
  \label{fig:phase}
\end{figure}

To elaborate on the phase diagram, let's look at the $S=3/2$ case. The density-matrix renormalization group calculation\cite{PhysRevLett.69.2863} is performed to obtain the energy of the rung singlet plus the spin-one, spin-two, and spin-three Haldane phases for different values of $J_{\perp}/J_1$ as shown in Fig.\ref{fig:phase}a. The spin-three Haldane state is the ground state when $J_{\perp}/J_1$ is smaller than a critical value $J_{\perp}/J_1 \approx 1.68$. When $J_{\perp}/J_1$ is larger than this critical value, the ground state is the rung singlet. 
We can further find the critical point that separates two distinct phases for other spin values. The critical point would finally move toward $J_{\perp}/J_1=2$ as $S$ goes to infinity. As a result, we draw a schematic phase diagram as shown in Fig.\ref{fig:phase}b where a quantum phase transition line separates the rung singlet and the spin-$2S$ Haldane phase.

\begin{figure}
  \includegraphics[width=8cm]{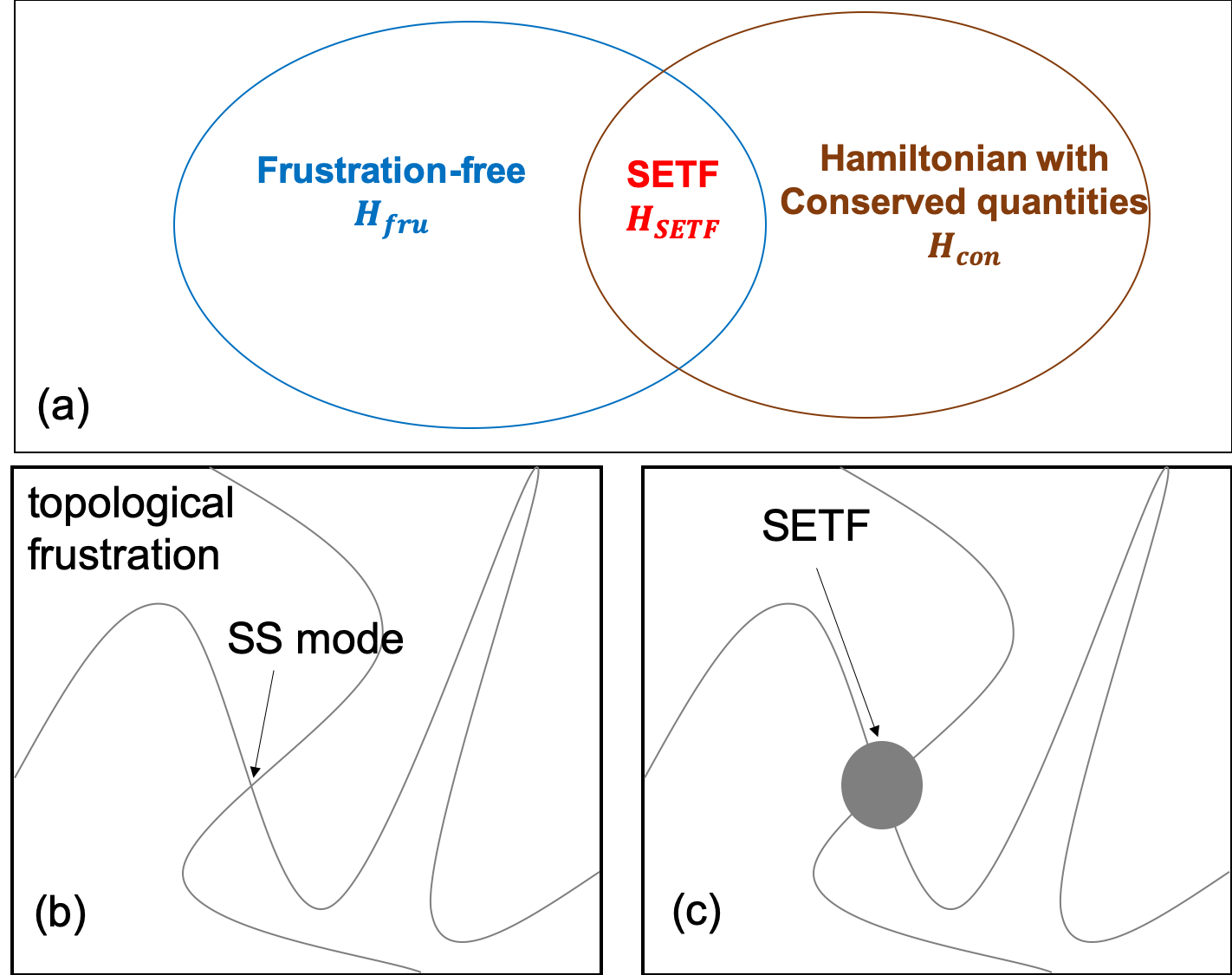}
  \caption{(a)Relations among symmetry-enriched topological frustration(SETF), frustration-free Hamiltonian, and the Hamiltonian with conserved quantities. (b)The topological space of zero modes with a self-stress(SS) mode (c)The topological space of zero modes with a regime where the zero modes have extra dimensions.}
  \label{fig:TF}
\end{figure}

The region of the phase diagram where $H_{ladder}$ can be placed in the form of $H_{fru}$ lies on the upper part of Fig.\ref{fig:phase}(b) where $J_{\perp}/J_1 \geq 2$ and labeled "classical frustration". But we also restricted parameters so that we can write $H_{ladder}$ as $H_{con}$, as discussed above. Thus there is an overlap between $H_{con}$ and $H_{fru}$ where the spin ladder is classically frustrated and has local conserved quantities (See Fig.\ref{fig:TF}a). In the overlap case, the Hamiltonian can always be written in a frustration-free form
\begin{equation}
    H_{SETF}=\frac{J_1}{2S(S+1)}\sum_{i}{(b_1\overrightarrow T_{i} + \overrightarrow b_2T_{i+1})^2}
\end{equation}
which we will turn out to be the regime where the SETF occurs.

\section{The fate of topological frustration}
Now we have enough ingredients to study the connections between topological frustration and quantum magnetism. Especially, we will begin on the classical side to understand topological frustration and then to see what the fate of this topological frustration would be after turning on quantum mechanics. 

To do so, we start with the purely classical problem of finding the ground state of $H_{fru}$ in which the zero modes are the configurations which satisfy a set of constraints $a_1\overrightarrow S_{i,1} + a_2\overrightarrow S_{i,2}+a_3\overrightarrow S_{i+1,1}+ a_4 \overrightarrow S_{i+1,2}=0$. Based on Maxwell's counting\cite{Maxwell1864}, since each rung of the ladder has four degrees of freedom and three constraints (in average) the zero mode has one remaining degree of freedom for each rung. For example, when $a_1/a_2=c a_3/a_4$ where $c$ is some constant, we can define a vector $\overrightarrow V_{i} = a_1 \overrightarrow S_{i,1} + a_2\overrightarrow S_{i,2}$ and rewrite the constraint as $\overrightarrow V_{i}+c\overrightarrow V_{i+1}=0$. In this case the system has a local zero mode at each rung in which two spins combined can rotate as $U(1)$ symmetry about the axis described by $\overrightarrow V_{i}$.

Maxwell’s count is, however, incomplete, as discussed by Kane and Lubensky to linear level for balls-and-springs models\cite{Kane2013}. For example, considering that we have $n$ degrees of freedom and $n-1$ constraints. In a generic case, the topological space of zero modes would look like a one-dimensional manifold except for some points where two curves (or more than two curves) intersect (See Fig.\ref{fig:TF}b). Those intersecting points are the places where self-stress modes appear and give an extra number of zero modes in a linear theory. 

In a full nonlinear problem, the topological space of zero modes can change dramatically due to certain symmetry that makes some constraints become redundant. This gives extra dimensions to zero modes, the phenomenon we call SETF (See Fig.\ref{fig:TF}c). In the Hamiltonian $H_{fru}$, for example, when $a_1=a_2$ and $a_3=a_4$, which corresponds to $J_{\parallel,1}=J_{\parallel,2}=J_{X,1}=J_{X,2}$, the configurations with two spins at the same rung pointing into opposite directions ($\overrightarrow S_{i,1}=-\overrightarrow S_{i,2}$) is a local zero mode in which two spins combined can rotate as $SU(2)$ symmetry which has two continuous degrees of freedom that is one more than Maxwell's counting. In this SETF regime, we can always rewrite $H_{fru}$ in a form of $H_{SETF}$ by defining $\overrightarrow T_i=\overrightarrow S_{i,1}+ \overrightarrow S_{i,2}$. Thus, the regime where SETF occurs is exactly the overlap between $H_{con}$ and $H_{fru}$.

\begin{figure}
  \includegraphics[width=8cm]{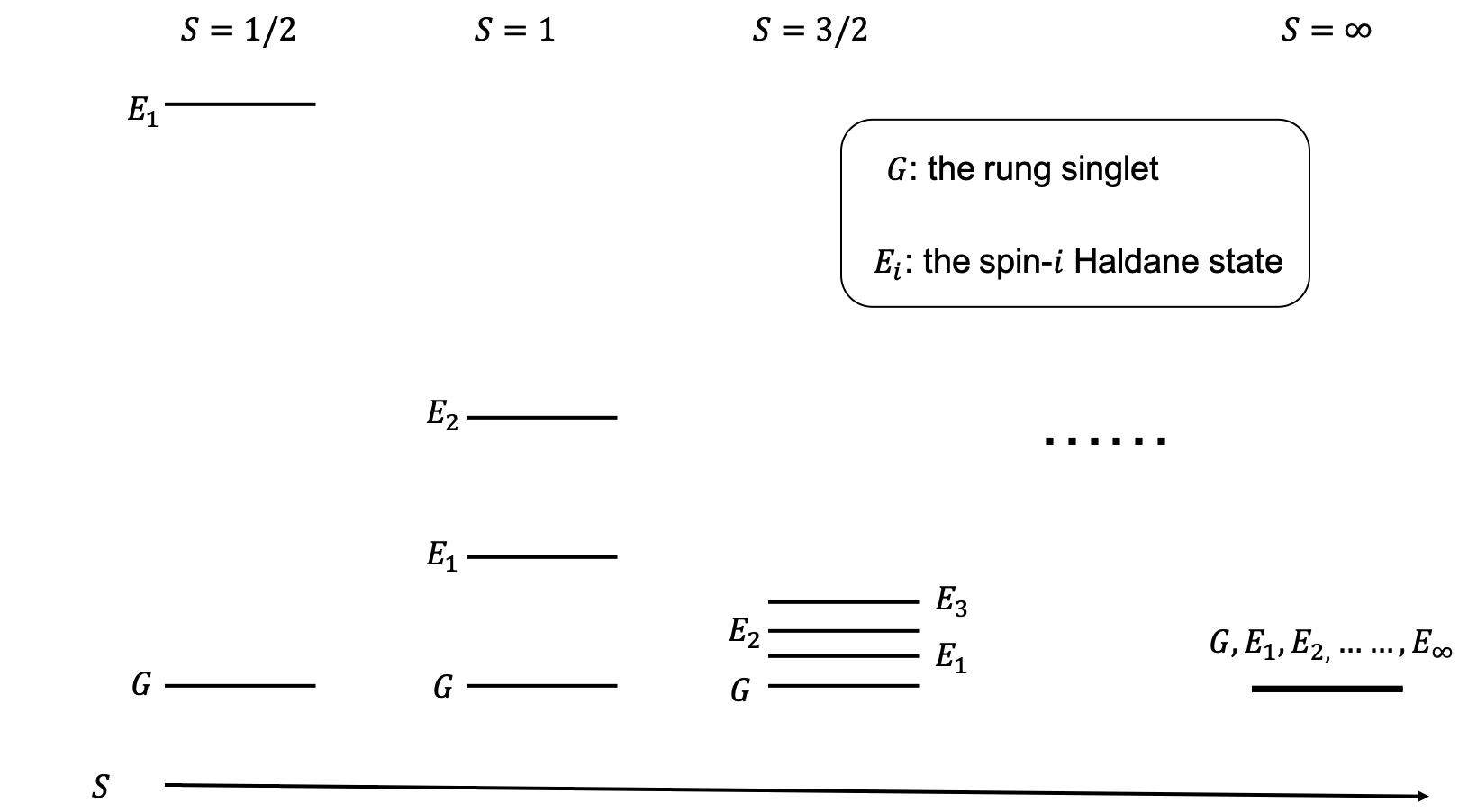}
  \caption{Low energy eigenstates at different spins for the spin ladder model.}
  \label{fig:state}
\end{figure}

To understand how this SETF is preserved from infinite $S$ to finite $S$, let's take a highly frustrated point $J_{\perp}/J_1=2$ for an example. The corresponding Hamiltonian is written as
\begin{equation}
    H_{2}=\frac{J_1}{2S(S+1)}\sum_{i}{(\overrightarrow T_{i,1}+ T_{i+1,1})^2}.
\end{equation}
As we move from infinite $S$ to finite $S$, the strict zero modes of the classical limit all get lifted by quantum fluctuations and we are left with a unique rung singlet ground state $G$ (Fig.\ref{fig:state}). But the SETF at finite but large $S$ is preserved as the existence of many very low energy excitations. We know this exactly by mapping them to the SPT spin-$n$ Haldane states $E_n$ whose topological properties are protected by the staggered swap symmetry (See Appendix A).

It turns out there is a simple argument that predicts the lifting of the classical zero modes by quantum fluctuations. The recently developed theory of nonlinear topological mechanics\cite{PhysRevLett.127.076802} identifies a topological invariant that protects the existence of classical zero modes by surface integrals over phase space. If these constraints only involve position variables, the surfaces are well defined both at the quantum and classical levels. So it could be the topology is preserved by quantum fluctuations and captured by a quantum version of nonlinear topological mechanics. However, in the present case, these surfaces are defined by the constraints in Eq. \ref{eq:constraint} that arise from angular momentum variables that involve position and momentum variables. So, upon quantizing the system, the surfaces cease to exist by the Heisenberg uncertainty principle and the topological invariant becomes undefined. As a result, it is not surprising the topology is lost in the finite $S$ model.

Though the zero modes are lifted at finite $S$, the staggered swap symmetry allows some of them to become SPT states. This symmetry allows us to rewrite the frustration-free Hamiltonian in terms of a new spin operator defined by a pair of spins ($\overrightarrow T_{\alpha}=\overrightarrow  S_b + \overrightarrow  S_c$), the resulting Hamiltonian would have a local conserved quantity $T_\alpha^2$. In classical theory, this symmetry demands one more degree of freedom than Maxwell's counting would predict.  The conservation of $T_\alpha^2$ groups the Hilbert space into different sectors each labeled by its eigenvalues $T_{\alpha}(T_{\alpha}+1)$. Each sector is characterized by its own Hamiltonian with its own "ground states" and set of excitations. Therefore, the classical SETF phase in the spin ladder has asymptotically low-energy topological eigenstates whose presence is the quantum manifestation of a combination of topological frustration and the staggered swap symmetry.

\section{Discussion and outlook}
Topological frustration is a classical phenomenon that, in principle, is able to study even in an unsolved frustrated spin system, but does not draw much attention due to the ignorance of the connections between classical frustration and quantum magnets. 
In particular, combined with the role of symmetry, the fate of SETF and how SETF is preserved from infinite spin to a finite spin points out a new direction to study unsolved frustrated spin systems such as kagome and pyrochlore antiferromagnets.

Below we discuss several future directions this research motivates: the potential observation of quantum scars, two-dimensional topological frustration systems with a similar set of conservation laws, and the potential use of tensor network methods to study large-$S$ frustrated magnets.

\subsection{Existence of quantum scars}

In the nearly SETF regime where the conservation laws are violated due to some small perturbation, we speculate the quantum scars could be observed. An isolated quantum system was believed to be thermalized in a way such that the system can be described by equilibrium statistical mechanics that we call the eigenstate thermalization hypothesis (ETH). However, some quantum systems were found disobeying the ETH\cite{PhysRevA.43.2046}. In particular, when there exist many conserved quantities in a quantum system the ETH is strongly violated which is essentially the case where SETF occurs. If we move slightly away from the SETF regime by adding some small perturbation, the system becomes weakly ETF breaking, and quantum scars might be observed\cite{PhysRevB.98.155134}.

To illustrate the idea, let's look at quantum spin ladders as a concrete example. For quantum spin ladders with the staggered swap symmetry, the ground states (the Haldane states) in different sectors are gapped to their excited states of the same sector. With a small perturbation, interactions can be introduced between two states with the same energy but in different sectors. For example, the ground states in two sectors $(|T_1|,|T_2|,|T_3|,|T_4|,......)=(0,1,0,1,......)$ and $(|T_1|,|T_2|,|T_3|,|T_4|,......)=(1,0,1,0,......)$ have the same energy but are non-interacting when staggered swap symmetry holds. With small perturbation that breaks the staggered swap symmetry, those two states can become interacting. In this case, if we prepare the ground state in the sector $(|T_1|,|T_2|,|T_3|,|T_4|,......)=(0,1,0,1,......)$ as an initial state, we speculate that similar to a quantum scarred eigenstates in a Rydberg atom chain\cite{NaturePhysics14,PhysRevB.98.155134}, an oscillation between the two states $(0,1,0,1,......)$ and $(1,0,1,0,......)$  might be observed. Similarly, other initial states could also lead to different patterns of quantum scars in the nearly SETF regime.

\subsection{Generalization to higher dimensions}

\begin{figure}
  \includegraphics[width=8cm]{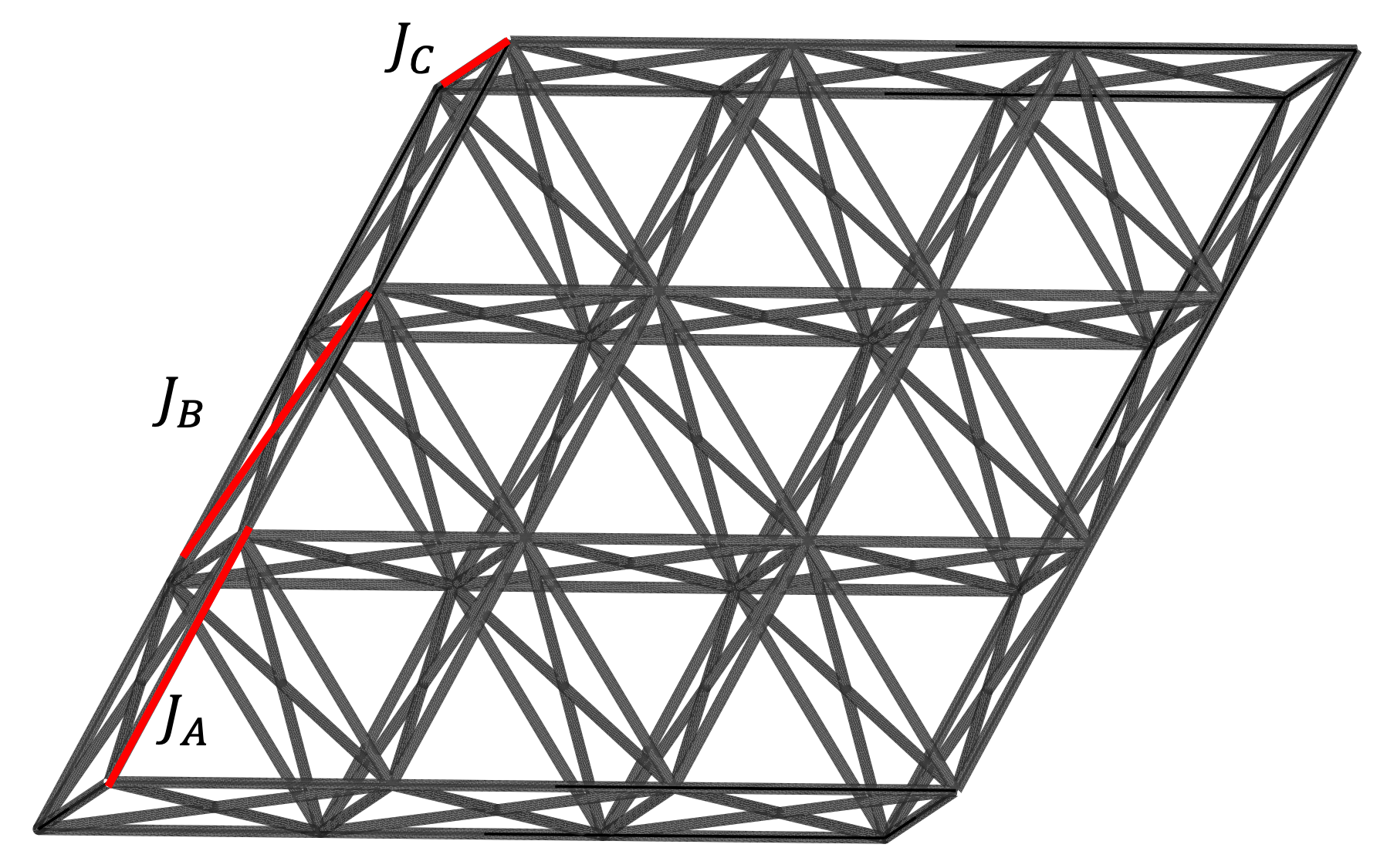}
  \caption{The bilayer triangular lattice model}
  \label{fig:bilayer}
\end{figure}

Based on the special role of $SU(2)$ symmetry in SETF, we design a bilayer triangular lattice model as shown in Fig.\ref{fig:bilayer}. The Hamiltonian is
\begin{equation}\label{eq::bilayer}
\begin{split}
    H_{bilayer}=\frac{J_{A}}{S(S+1)}\sum_{i,j,m}(\overrightarrow S_{i,j,m} \cdot \overrightarrow S_{i+1,j,m}\\ 
    + \overrightarrow S_{i,j,m} \cdot \overrightarrow S_{i,j+1,m}
    +\overrightarrow S_{i+1,j,m} \cdot \overrightarrow S_{i,j+1,m})\\
    +\frac{J_{B}}{S(S+1)}\sum_{i,j}(\overrightarrow S_{i,j,1} \cdot \overrightarrow S_{i+1,j,2}
    + \overrightarrow S_{i,j,1} \cdot \overrightarrow S_{i,j+1,2}\\
    +\overrightarrow S_{i+1,j,1} \cdot \overrightarrow S_{i,j+1,2}
    +\overrightarrow S_{i,j,2} \cdot \overrightarrow S_{i+1,j,1}\\ 
    + \overrightarrow S_{i,j,2} \cdot \overrightarrow S_{i,j+1,1}
    +\overrightarrow S_{i+1,j,2} \cdot \overrightarrow S_{i,j+1,1})\\
    +\frac{J_{C}}{S(S+1)}\sum_{i,j}\overrightarrow S_{i,j,1} \cdot \overrightarrow S_{i,j,2}
\end{split}
\end{equation}
where $\overrightarrow S_{i,j,m}$ is the spin operator at the site $(i,j,m)$, and $J_{A}$, $J_{B}$, and $J_{C}$ are the antiferromagnetic coupling strengths depicted in Fig.\ref{fig:bilayer}.

Topological frustration occurs when $J_C=2J_B$. Under this condition, the Hamiltonian can be written in a frustration-free form
\begin{equation}
\begin{split}
    H_{bilayer(fru)}=\frac{J_B}{2S(S+1)}\sum_{i,j}(\overrightarrow S_{i,j,1} + \frac{J_A}{J_B} \overrightarrow S_{i,j,2}\\
    \overrightarrow S_{i+1,j,1} + \frac{J_A}{J_B} \overrightarrow S_{i+1,j,2}\\
    +\overrightarrow S_{i,j+1,1}
    + \frac{J_A}{J_B} \overrightarrow S_{i,j+1,2})^2
\end{split}
\end{equation}
which can then be understood as a constraint problem in the classical limit. Similar to the spin ladder model, each vertex of triangles has two spins and thus four degrees of freedom. In average, there are three constraint for each vertex, so the zero mode has one remaining degree of freedom on each vertex shared by two spins. 

Topological frustration is enriched when $J_A/J_B=1$. At this point, we can define a new set of spin operators $\overrightarrow T_{i,j} = \overrightarrow S_{i,j,1}+\overrightarrow S_{i,j,2}$, and rewrite the Hamiltonian as
\begin{equation}
\begin{split}
    H_{bilayer(SETF)}=\frac{J_B}{2S(S+1)}\sum_{i,j}(\overrightarrow T_{i,j}
    + \overrightarrow T_{i+1,j}
    +\overrightarrow T_{i,j+1})^2.
\end{split}
\end{equation}
Configurations with two spins on the same vertex pointing into opposite directions ($\overrightarrow S_{i,j,1}=-\overrightarrow S_{i,j,2}$) are the zero modes with two continuous degrees of freedom on each vertex. Therefore, the same form of SETF occurs in this bilayer triangular lattice model. 

\begin{figure}
  \includegraphics[width=8cm]{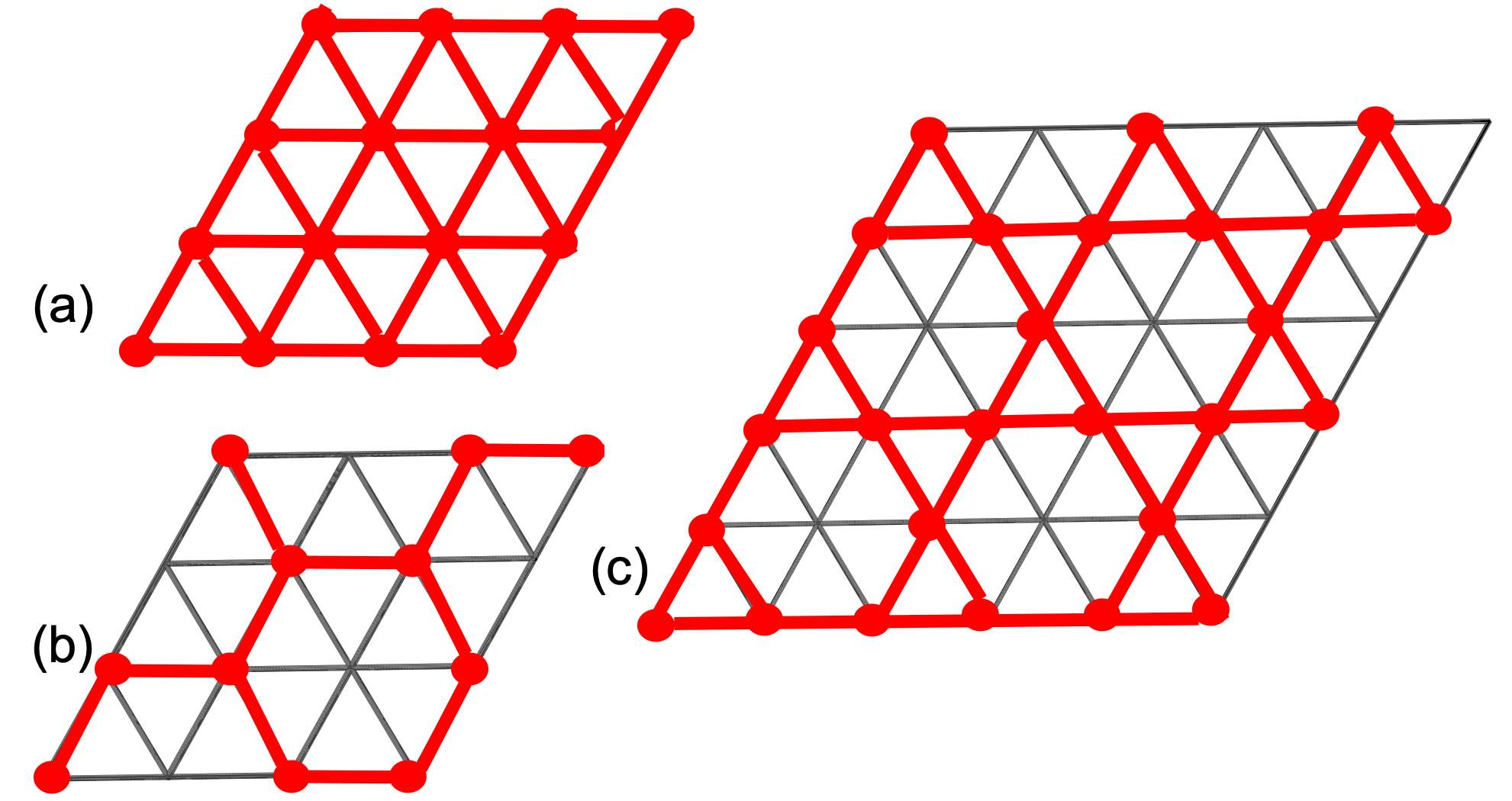}
  \caption{(a)A triangular lattice model (b)A honeycomb lattice model (c)A kagome lattice model}
  \label{fig:3examples}
\end{figure}

As we go from infinite $S$ to a finite value of $S$, each spin operator $\overrightarrow T_{i,j}$ gives a local conserved quantity $T_{i,j}(T_{i,j}+1)$. A set of low energy eigenstates of the bilayer triangular lattice model is consisted of the ground states from different sectors defining by infinite many conserved quantities $T_{i,j}(T_{i,j}+1)$. Some of those eigenstates come from well-known models that have been studied by previous works\cite{PhysRevB.97.245146,PhysRevB.102.121102,PhysRevB.91.100407}. For example, when $T_{i,j}(T_{i,j}+1)=2$ for all $i,j$, $H_{bilayer(SETF)}$ is reduced to spin-one Heisenberg triangular lattice model [See Fig.\ref{fig:3examples}(a)] which has the 120 degree magnetically ordered ground state\cite{PhysRevB.97.245146}. We can also make some $T_{i,j}(T_{i,j}+1)=0$ and some other $T_{i,j}(T_{i,j}+1)=2$ to obtain spin-one Heisenberg honeycomb and kagome lattice models as shown in Fig.\ref{fig:3examples}(b) and (c). The ground state of 
spin-one Heisenberg honeycomb lattice model has been found to be a Neel state while a possible candidate for the ground state of spin-one Heisenberg kagome lattice model is the hexagon singlet solid\cite{PhysRevB.102.121102,PhysRevB.91.100407}.
From the above analysis, we can study the spectra features with the benefit of being able to calculate some topological eigenstates from a simplified model based on a set of conserved quantities. Moreover, we can further infer the spectra features for some unsolved model such as kagome and pyrochlore antiferromagnets by understanding the fate of SETF.

\subsection{Tensor network methods}

For a generic model, understanding the fate of SETF and how it is preserved in the quantum limit relies on numerical simulations. In particular, tensor network methods have been shown powerful to handle low-dimensional frustrated spin systems. Especially in one-dimensional systems, tensor network methods have been well developed from the matrix product state\cite{PhysRevLett.69.2863}. Nevertheless, the challenge significantly grows in dealing with two-dimensional systems because of the need for large size tensors to obtain a wavefunction with acceptable accuracy. To overcome the challenge several algorithms such as the projected entangled pair states, the infinite projected entangled pair states, and infinite projected entangled simplex states have been developed to reduce fitting parameters of tensors based on symmetry\cite{PhysRevA.70.060302,PhysRevLett.101.250602,PhysRevX.4.011025}. The idea behind those algorithms implies that even with large size tensors one can still use only a small number of parameters by appropriately imposing structures on tensors. In other words, as we move to large spin cases, although the size of tensors may increase, it is possible to use fewer parameters to construct a tensor network representation of a wavefunction with the same accuracy as that in the spin-half or one case. 

To see whether we can study the large spin cases in the spin ladder model with achievable computational resources, we first notice that mutual information $I$ in the classical limit can be the analogy to entanglement entropy $S_q$. For the rung singlet phase, since knowing the directions of two spins at a certain rung does not give us any information on the directions of spins at other rungs, the mutual information is zero, and so is the entanglement entropy. On the other hand, in the case of the Haldane phase, classically, once we know the direction of spins at a certain rung, the directions of the rest of the spins can be completely determined. Thus, both the mutual information and entanglement entropy are nonzero. The mutual information can be computed as follows. The (classically) entropy $S_c$ for a chain with any size is always $lnN$ where $N$ is the number of states (assume a uniform grid $N$ on a sphere) for an individual spin. Now if the system is divided into subsystem $A$ and subsystem $B$, the mutual information between them would be 
\begin{equation}
I = S_c(A)+S_c(B)-S_c(A+B) = lnN+lnN-lnN=lnN.
\label{eq:mutual}
\end{equation}
From the analog between mutual information and entanglement entropy, Eq.\ref{eq:mutual} implies that the entanglement entropy has an asymptotic function $ln(2S+1)$ as $S$ goes infinity where $2S+1$ is the degrees of freedom for a quantum spin-$S$. Thus a tensor network representation for a large spin-$S$ Haldane state can be constructed by tensors with virtual bond dimension to the order of $2S+1$

\begin{figure}
  \includegraphics[width=8cm]{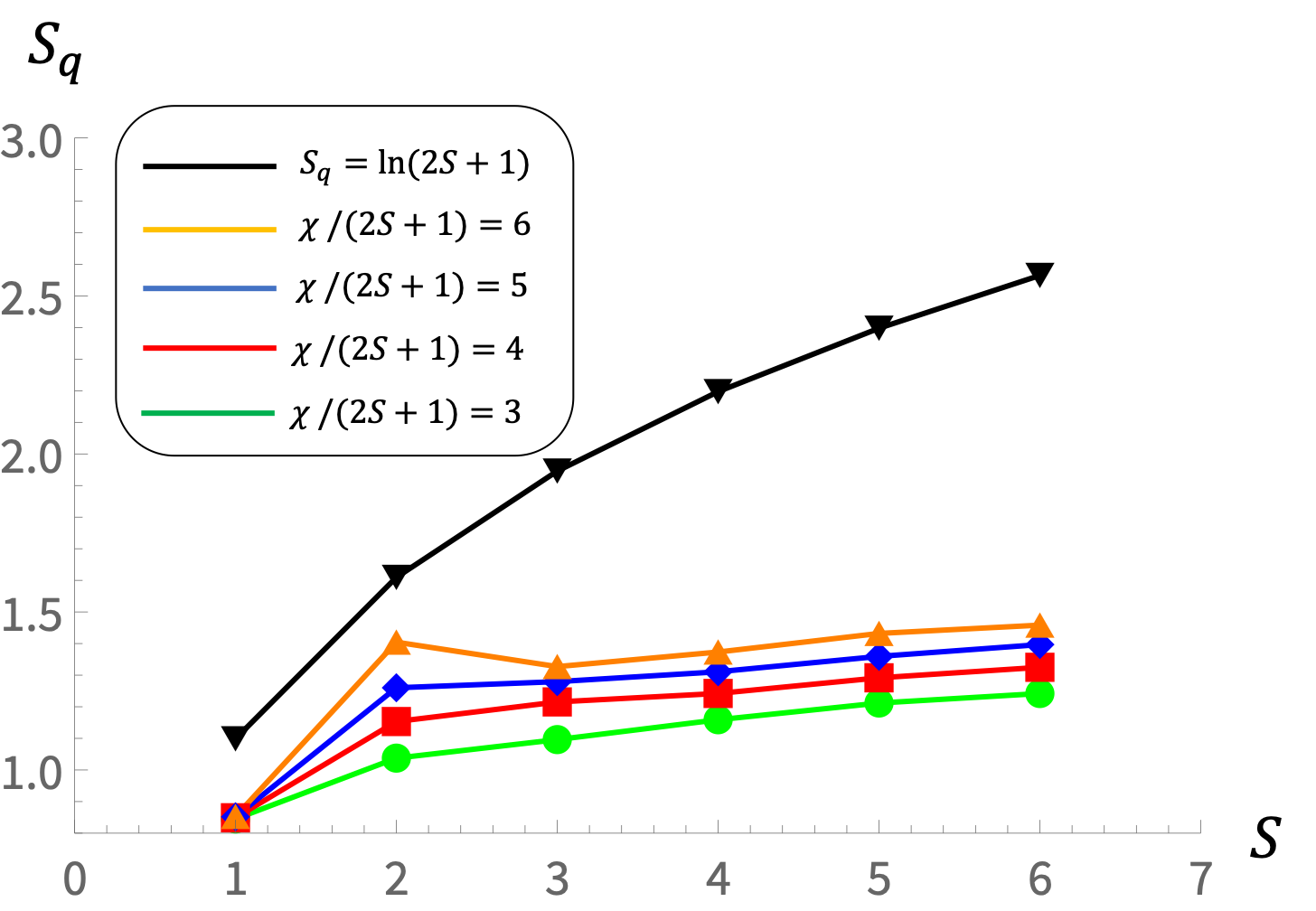}
  \caption{The density-matrix renormalization group calculation for the entanglement entropy as a function of $S$ for different virtual bond dimension $\chi/(2S+1)$ of the Heisenberg spin chain (50 spins).}
  \label{fig:entropy}
\end{figure}

To further confirm our claim, we perform the density-matrix renormalization group calculation for the Heisenberg spin chain as shown in Fig.\ref{fig:entropy}. Here we compare different spin cases with the same ratio of the virtual bond dimension $\chi$ to $2S+1$. The factor $2S+1$ is able to be factored out by using appropriate symmetric tensors because it comes from the global $SU(2)$ symmetry. With different values of $\chi/(2S+1)$, the entanglement entropy is always bounded by $ln(2S+1)$. Moreover, the entanglement entropy only increases slightly with the increasing $S$. As a result, we conclude that in the frustrated spin ladder model, the tensor network method can be used to study the large spin regime which would give us a better understanding of SETF.


\section{Acknowledgements}
We thank Nic Shannon (OIST) for useful discussions. This material is based upon work supported by the National Science Foundation under Grant No. OAC-1940260.

\bibliographystyle{unsrt}
\bibliography{bib}

\appendix

\section{}

We study the spin-1/2 ladder model with 6 rungs to see what happens to the Haldane spin-one SPT eigenstates after breaking the staggered swap symmetry ($J_{\parallel,1}=J_{\parallel,2}=J_{X,1}=J_{X,2}$).

\begin{figure}
  \includegraphics[width=8cm]{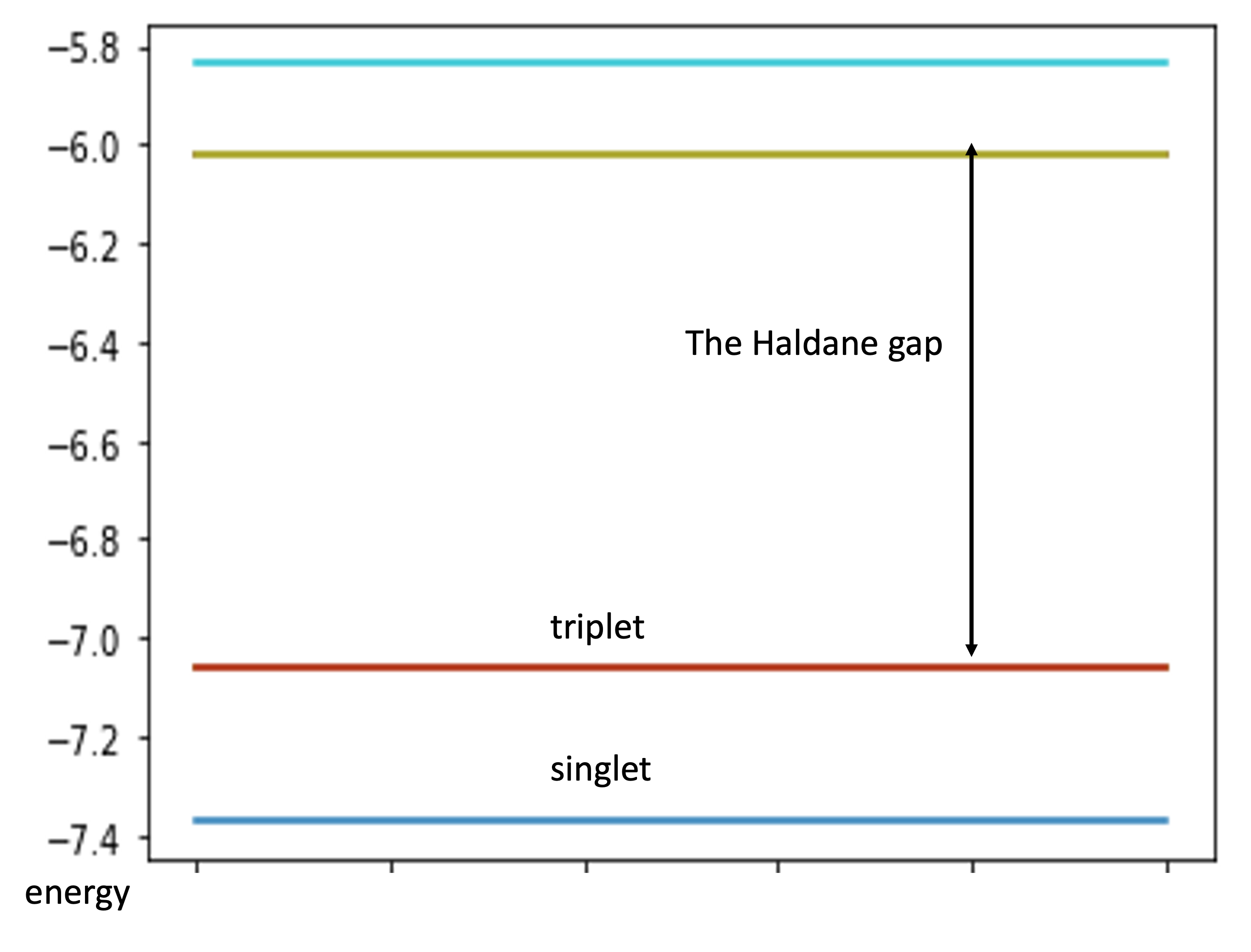}
  \caption{The spectrum of the antiferromagnetic Heisenberg spin-one chain with 6 sites.}
  \label{fig:HaldaneGap}
\end{figure}

\begin{figure}
  \includegraphics[width=8cm]{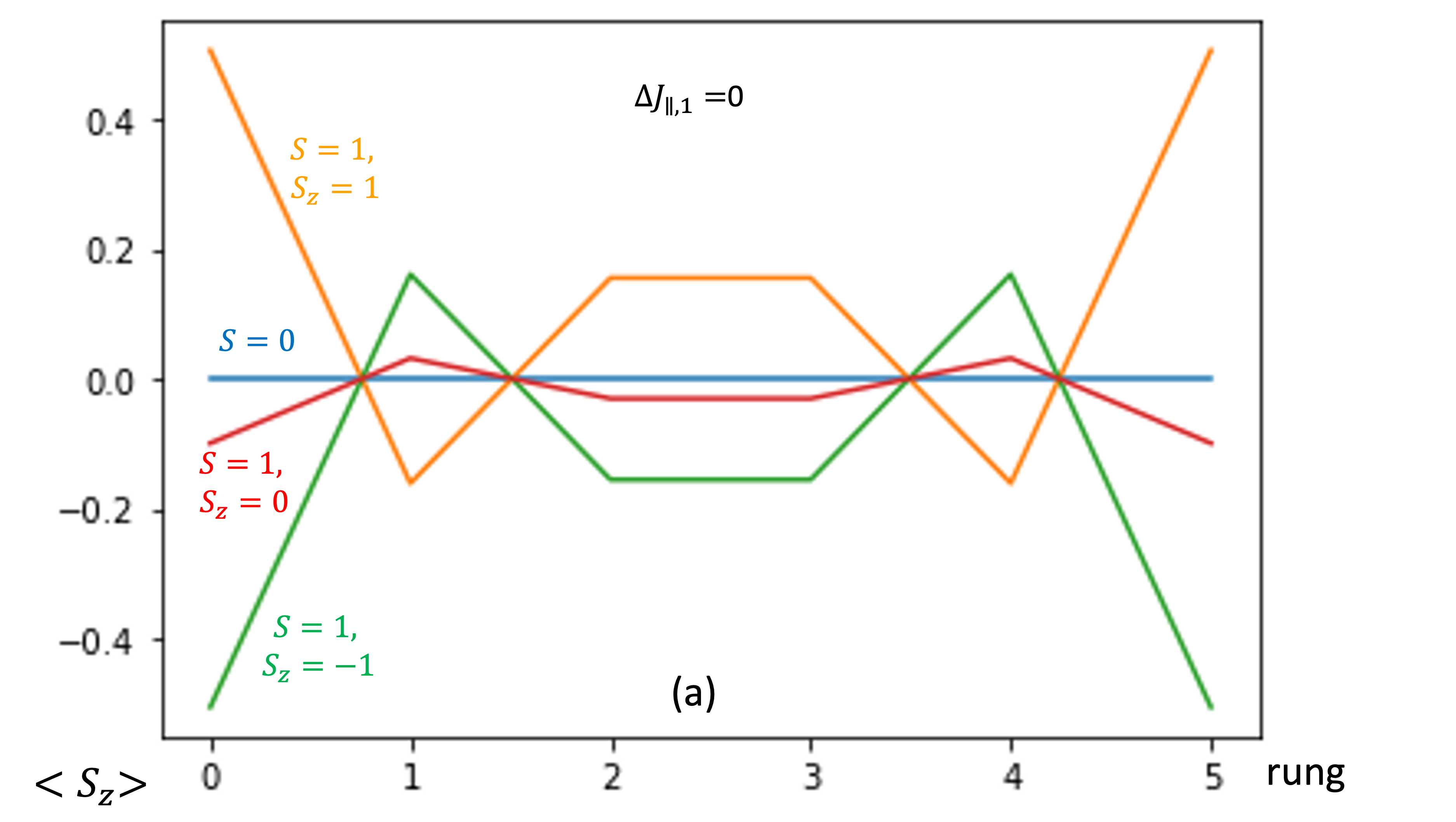}
  \includegraphics[width=8cm]{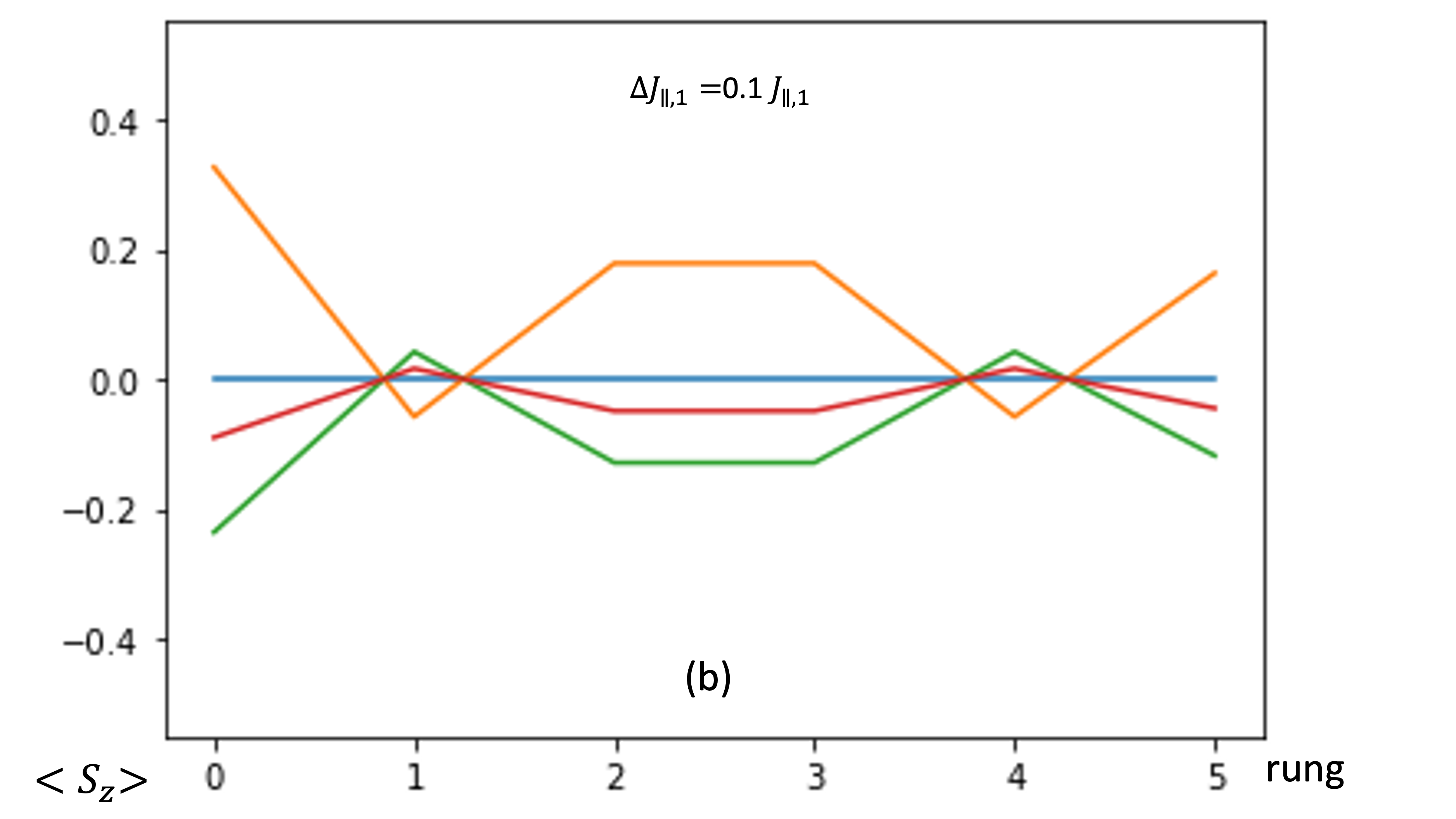}
  \includegraphics[width=8cm]{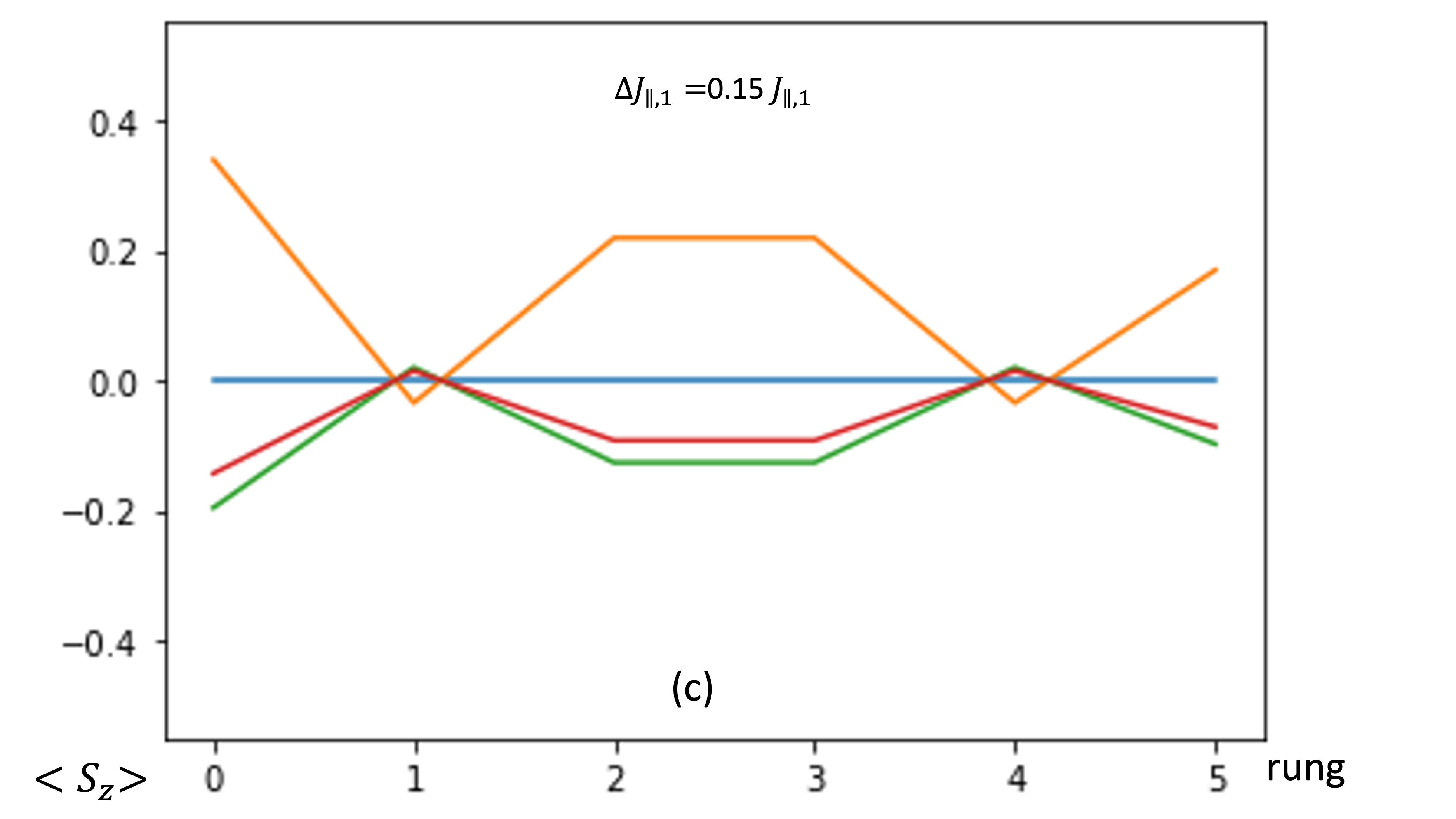}
  \caption{(a)The expectation value of $S_z$ on each rung for the $S=0$ singlet and $S=1$ triplet. (b) and (c) show the evolution of $<S_z>$ after breaking the staggered swap symmetry.}
  \label{fig:evolution}
\end{figure}

In the antiferromagnetic Heisenberg spin-one chain model, the lowest 4 eigenstates are separated by a gap from the other states as shown in Fig\ref{fig:HaldaneGap}. Those four states are $S=0$ singlet and $S=1$ triplet. The excitation ($S=1,S_z=-1,1$) of the ground state ($S=0$) is an edge state that has two spin-half particles separately at two edges where the expectation value of $S_z$ is roughly $0.5$ as show in Fig.\ref{fig:evolution}(a).

Now we start with the ladder model at the highly frustrated point $J_{\perp}=2J_{\parallel,1}=2J_{\parallel,2}=2J_{X,1}=2J_{X,2}$ and identify the corresponding $S=0$ singlet and $S=1$ triplet are the 54th, 61st, 62nd and 63rd excited states.

When we change $J_{\parallel,1}$ slightly without having any level crossing, $S_z$ of those four states change as Fig.\ref{fig:evolution}(b) and (c). When $\Delta J_{\parallel,1} = 0.01 J_{\parallel,1}$, from the expectation values of $S_z$ on different rungs we can see the states $S=1,S_z=-1,1$ evolve into the bulk. In these two states, $<S_z>$ at two edges decreases and $<S_z>$ of the middle rungs increases becoming comparable to $<S_z>$ at two edges[See Fig.\ref{fig:evolution}(b)]. When we increase $\Delta J_{\parallel,1}$ further to $0.01 J_{\parallel,1}$ as shown in Fig.\ref{fig:evolution}(c) the state $S=1,S_z=-1$ significantly mixed with the state $S=1,S_z=0$. From the evolution of $<S_z>$ in Fig.\ref{fig:entropy}(a)-(c), we see that the edge states are not robust with the deformation (change of $J_{\parallel}$). In other words, these Haldane eigenstates are protected by the staggered swap symmetry ($J_{\parallel,1}=J_{\parallel,2}=J_{X,1}=J_{X,2}$) that is also the symmetry enriches the topological frustration in the classical limit.

\end{document}